\documentclass[letterpaper, 10 pt, conference]{ieeeconf}  
\usepackage{blindtext,graphicx,booktabs,amsmath}
\usepackage{graphicx,amsmath,cite}
\usepackage{wrapfig}
\usepackage{lipsum}
\usepackage{textcomp}
\usepackage{mathtools}
\usepackage{fancyhdr}
\usepackage{graphics}
\usepackage{setspace}
\usepackage{epstopdf}
\usepackage{multirow}
\usepackage{setspace}
\usepackage{booktabs}
\usepackage{array}
\usepackage{amssymb}
\usepackage{verbatim}
\usepackage{cleveref}
\usepackage{color}
\usepackage{relsize}
\usepackage{caption}
\usepackage{subfigure}
\IEEEoverridecommandlockouts

\title{\LARGE \bf Evaluation of the Energy Efficiency in a Mixed Traffic with Automated Vehicles and Human Controlled Vehicles}

\author{Xun~Gong$^1$, Yaohui Guo$^1$, Yiheng Feng$^2$, Jing Sun$^3$, Ding Zhao$^1$
\thanks{This work is funded by the United States Department of Energy (DOE), ARPA-E NEXTCAR program.}
\thanks{$^{1}$ X. Gong, Y. H. Guo and D. Zhao are with the department of Mechanical Engineering, University of Michigan, Ann Arbor, MI, 48109, USA.}
\thanks{$^{2}$Y. H. Feng is with the Transportation Research Institute, University of Michigan, Ann Arbor, MI, 48109, USA.}
\thanks{$^{3}$J. Sun is with the Department of Naval Architecture and Marine Engineering, University of Michigan, Ann Arbor, MI, 48109, USA.}
\thanks{$^{*}$Corresponding author. E-mail: zhaoding@umich.edu (D. Zhao)}
}

\begin{document}
\maketitle

\begin{abstract}
The energy efficiency of Connected and Automated Vehicles (CAVs) is significantly influenced by surrounding road users.
This paper presents the evaluation of energy efficiency of CAVs in a mixed traffic interacted with human controlled vehicles. To simulate the interaction between the CAVs and the cut-in vehicles controlled by human drivers near the intersection, a lane changing model is proposed to emulate the politeness and patience characteristics of the human driver. The proposed lane changing model is then calibrated based on over 100,000 naturalistic lane changing events
collected by the University of Michigan Safety Pilot Model
Deployment Program. A case study on simulation of the cut-in scenario is carried out to demonstrate the human driver's lane changing sensitivity under different driving trajectories of a frontal CAV and the influence on the energy consumption of the CAV due to the cut-in vehicle is evaluated. The simulation results indicate that the fuel economy of the CAV can be substantially improved if its surrounding cut-in vehicles can be well handled.
\end{abstract}


\IEEEpeerreviewmaketitle

\section{Introduction}
In recent years, the dedicated short range communication (DSRC) for vehicle-to-vehicle (V2V) and vehicle-to-infrastructure (V2I/V2X) has become more integrated and ubiquitous, leading to the emergence of connected and automated vehicles (CAVs) with enhanced safety and improved fuel economy~\cite{Sciarretta2015}.
The Eco-approaching and departure (Eco-AD) is the core technology to improve fuel economy of CAVs in urban driving~\cite{Barth2009}. In the early stage of the development of CAVs (or the stage of low penetration rate), the Eco-AD strategy and the associated energy efficiency will depend on the operation of surrounding road users.

\begin{figure}[h]
\renewcommand{\captionfont}{\small}
\centering\includegraphics[width=1\linewidth]{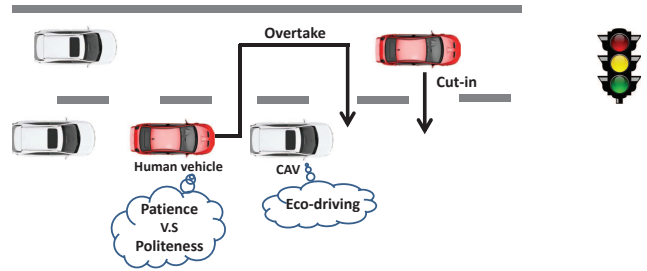}
\caption{Illustration of the mixed urban traffic scenario involving CAVs and cut-in vehicles near the intersection.}
\label{fig:task1s2:motivation}
\end{figure}

In a mixed traffic condition near the intersection as shown in Fig.~\ref{fig:task1s2:motivation}, if a CAV is driving under a predefined trajectory with a relative slow speed aiming to obtain the optimum fuel economy, the following human drivers who expect a higher speed may lose their patience and decide to change the lane, overtake the CAV and then  cut in back. At the same time, the vehicles in the neighbor lane will also have the chance to cut in if the condition is appropriate. Due to the cut-in vehicles, the CAVs may lose the chance to pass the intersection and the expected fuel economy may not be guaranteed from intersection to intersection. Therefore, understanding the factors which influence human drivers' lane change behaviors and how CAVs will interact with the human controlled vehicles plays an important role in the evaluation of energy efficiency of CAVs.


In the last two decades, many studies have been published on lane changing models of human drivers. Early on, Gipps~\cite{Division1986} proposed a deterministic lane changing model concept based on gap-acceptance, in which a driver's behavior is governed by two basic considerations: maintaining a desired speed or being in the correct lane for an intended turning maneuver. In~\cite{Ahmed1999}, a probabilistic model was developed to describe lane changing decisions based on discrete choices. In~\cite{Toledo2007} and~\cite{Toledo2009}, Toledo, et al. developed an integrated model which allows to combine tactical and operational lane changing decisions by using the maximum likelihood estimation technique. In~\cite{Kesting2007}, by introducing the concept of drivers' politeness, Kesting proposed a novel acceleration-based logic model ('minimizing overall braking induced by lane changes', MOBIL) accounting for the anticipated advantages and disadvantages of a potential lane change. 
In~\cite{Cao2017}, an optimization-based lane changing model was formulated to determine the optimal position at which an instruction to change lanes was given through automotive navigation systems. In~\cite{Rios-Torres2017}, an optimal algorithm was developed for the coordination of CAVs' lane change behaviors at merging zones and the fuel consumption was evaluated through simulation. Recently, artificial intelligence based methodologies such as Neural Network~\cite{Hunt1994}, fuzzy logic~\cite{Wu2003} and decision tree~\cite{Motamedidehkordi2017} etc., have also attracted the attentions of researchers. In~\cite{Moridpour2010}, comprehensive overviews of lane changing models were presented.

While previous studies on lane changing model have provided various applications in safety evaluation~\cite{Jula2000}, route planning~\cite{Cao2017} and traffic management~\cite{Hidas2005}, etc., little attention has been paid on energy-oriented applications.
Motivated by the needs of energy evaluation in mixed traffic conditions, this paper aims to develop a microscopic lane changing model of human controlled vehicles and evaluate the energy efficiency of the influenced CAVs in the cut-in scenario. Particularly, inspired by MOBIL model as published in~\cite{Kesting2007}, a stochastic lane changing model extended with the patience factor is calibrated based on naturalistic driving data. The proposed model is able to emulate the politeness and patience characteristics of human drivers. Additionally, a standard Eco-AD algorithm is applied for CAV to generate velocity trajectory without considering the cut-in vehicles. The simulation is conducted to investigate the sensitivity of human driver's lane changing behavior under different velocity trajectories of a CAV. The corresponding energy efficiency of the CAV is evaluated.  

The outline of the paper is as follows: In Section \ref{section:Humanmodeling}, a stochastic lane changing model of the human driver is proposed. Then, the Eco-trajectory generation of the CAV is introduced in Section~\ref{section:CAVmodel}. Section~\ref{section:simulation} gives the simulation results of the lane changing sensitivity and the energy evaluation. Section~\ref{section:conclusion} summarizes the primary findings.

\section{Lane changing model of human drivers}\label{section:Humanmodeling}
\begin{figure}[htp!]
\renewcommand{\captionfont}{\small}
\centering\includegraphics[width=0.96\linewidth]{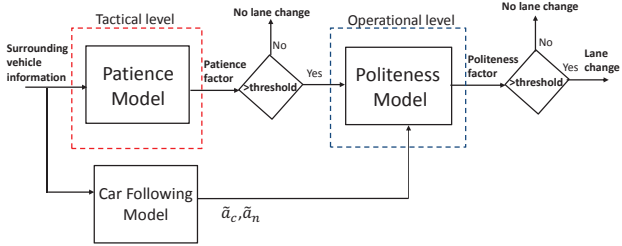}
\caption{The schematic diagram of the lane changing model.}
\label{fig:task1s2:lanechagediagram}
\end{figure}

In this section, a lane changing model of human drivers is developed. Two explanatory factors regarding the politeness and patience which affect drivers' lane changing decision are considered. Fig.~\ref{fig:task1s2:lanechagediagram} demonstrates the schematic diagram of the proposed lane changing model.
The lane change decision process is divided into the tactical level and the operational level. The patience model is executed in the tactical level which decides whether to pass a slow preceding vehicle or maintain the current speed according to the drivers' tolerance. In the operational level, the politeness model decides whether it is appropriate to execute a lane change maneuver according to the surrounding traffic condition accounting for drivers' courtesy. A specific car following model is applied to simulate the longitudinal driving behavior of the human controlled vehicle and provide the acceleration information for the politeness model. Details of the modeling procedure are introduced in the following subsections.

\subsection{Data collection}
In this study, the naturalistic database of the University of Michigan Safety Pilot Model Deployment (SPMD) program is utilized for the lane changing model development. The SPMD database is one of the largest public databases over the world that recorded the naturalistic driving over 34.9 million miles of 2,842 equipped vehicles in Ann Arbor, Michigan. 
Specifically, in SPMD database, there are 98 sedans equipped with the data acquisition system of MobilEye which enables to measure and record the relative speed, relative range, and the road curvature between the host vehicle and the preceding vehicle. Moreover, the speed, braking pedal and throttle angle were recorded from the CAN-bus of each host vehicle. 
The sampling frequency of the data was $10 Hz$.
\subsection{Politeness factor modeling}
\begin{figure}[htp!]
\renewcommand{\captionfont}{\small}
\centering\includegraphics[width=0.96\linewidth]{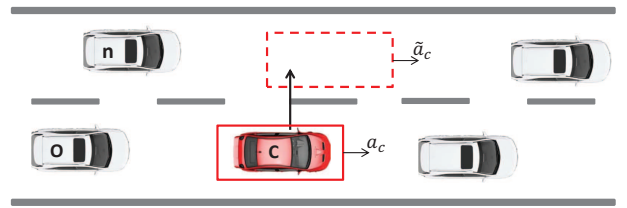}
\caption{Sketch of the nearest neighbors of a central human controlled vehicle c considering a lane change to the left lane.}
\label{fig:task1:Sketch}
\end{figure}

The lane changing scenario of human vehicles involving surrounding vehicles is shown in Fig.~\ref{fig:task1:Sketch}. In this paper, the idea of the lane changing model is inspired by the MOBIL concept in~\cite{Kesting2007} considering the two main advantages: (1) the lane changing decision process is significantly simplified; (2) the variables in the model are easily calculated with an underlying microscopic longitudinal traffic model, which enables the model to be easily integrated with a standard car following model. 

The desirability of the lane change rule is given based on the measurement of single-lane accelerations which indicate the anticipated advantages and disadvantages of a potential lane change by
\begin{equation}
\label{eq:task1:mobilmodelacc}
\underbrace{\tilde{a}_c-a_c}_{\text{driver}}+p(\underbrace{\tilde{a}_n-a_n}_{\text{new follower}}+\underbrace{\tilde{a}_o-a_o}_{\text{old follower}})>\Delta a_{th},
\end{equation}
where $p$ denotes the politeness factor and $\Delta a_{th}$ denotes the lane changing threshold, $\tilde{a}_{i}$ denotes the corresponding acceleration of vehicle $i$ after the lane changing and $a_i$ denotes the corresponding acceleration of vehicle $i$ before the lane changing. 

Since the disadvantage of the surrounding drivers and the advantage of the lane-changing driver are balanced via the politeness factor, the motivations for lane changing can be varied from purely egoistic to more cooperative behavior by adjusting the politeness factor. The rule of (\ref{eq:task1:mobilmodelacc}) indicates that lane changing execution considers the surrounding vehicles instead of the only successor in the target lane. Lane change should be only executed when the sum of acceleration of all involved vehicles is increased which corresponds to the concept of 'minimizing overall braking induced by lane changes' (MOBIL).

\begin{figure}[htp!]
\renewcommand{\captionfont}{\small}
\centering\includegraphics[width=0.88\linewidth]{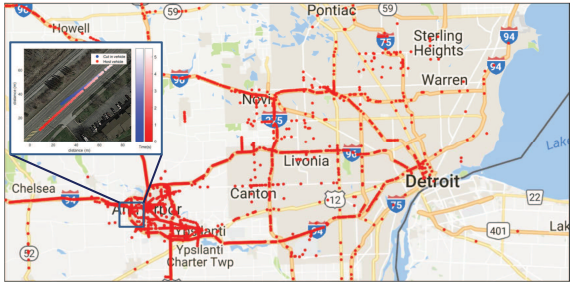}
\caption{Extracted lane change events distribution in Michigan area}
\label{fig:task1:eventdistribution}
\end{figure}

In this paper, the politeness factor $p$ is calibrated by using the SPMD database.
To achieve high query quality of the lane change records from the database, only lane change events on the straight lane are considered and extracted. Specifically, the following query criterion is applied to extract lane changing events/trajectories observed by host vehicles:
\begin{itemize}
  \item The host vehicle is not changing its lane;
  \item The frontal vehicle is cutting in;
  \item The lane is straight.
\end{itemize}
Consequently, $31,654$ qualified lane change events are successfully queried from the database. Fig.~\ref{fig:task1:eventdistribution} shows the distribution/locations of the identified lane changing events and the trajectory of one specific lane changing event in Michigan area. As the acceleration information is not directly available from the measurements, a specific differentiator filter is utilized to get the smooth acceleration applied in (\ref{eq:task1:mobilmodelacc}). 

Due to the limitation of the database that the measurement of the 'old follower' is not available, in current version of the lane changing model, we only consider the 'new follower' will interact with the lane changing vehicle and assume the acceleration of the 'old follower' is constant during the lane changing. Thus the politeness model is modified as
\begin{equation}
\label{eq:task1:modifiedmobilmodelacc}
\underbrace{\tilde{a}_c-a_c}_{\text{driver}}+p(\underbrace{\tilde{a}_n-a_n}_{\text{new follower}})>0,
\end{equation}
where the threshold $\Delta a_{th}$ is set as 0. 

\begin{figure}[h]
\renewcommand{\captionfont}{\small}
\centering\includegraphics[width=0.82\linewidth]{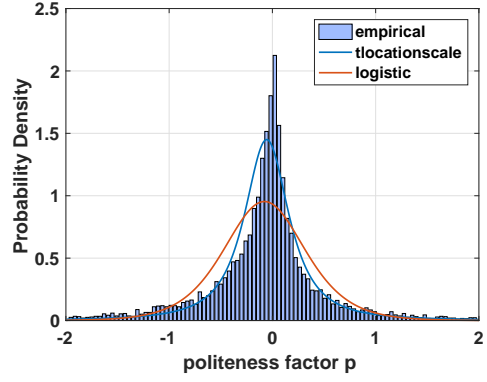}
\caption{Probability density distribution of the politeness factor $p$.}
\label{fig:task1:proacc}
\end{figure}

The probability density distribution of politeness factor $p$ is shown in Fig.~\ref{fig:task1:proacc}. It indicates that most of the drivers tend to perform with 'median' politeness ($p=[-1,1]$) when they execute lane change. The extremely egoistic ($p=-2$) and cooperative ($p=2$) drivers are as the minority. In order to further calibrate the distribution, specific numerical functions are used to fit the distribution. The \textit{t location-scale distribution} is the the closest numerical approximation to the actual distribution.

\subsection{Patience factor modeling}
The patience factor is another important factor that influences the human drivers' lane changing motivation. In this paper, the patience factor is modeled to determine whether or not an overtaking is willing to execute when the driver is following a vehicle with a relatively slow speed. The model formulation is given as follows depending on the summation of the velocity loss,

\begin{equation}
\label{eq:task1s2:patientmodel}
\sum_{N_{t1}}^{N_{t2}} (V_{des}-V_h)>\alpha_{pa},
\end{equation}
where $N_{t1}$ denotes sampling point number at the time when the human vehicle begins to follow a slow preceding vehicle and $N_{t2}$ denotes the sampling point number at the time when the human vehicle starts to execute lane changing, $V_{des}$ and $V_h$ denote the driver's desired velocity and actual velocity respectively, $\alpha_{pa}$  is the patience factor that reflects the human driver's tolerance for the velocity loss due to the slow preceding vehicle. The value of the driver's desired speed $V_{des}$ can be extracted from each lane change event in the naturalistic database. In this paper, we select $V_{des}$ as the vehicle's maximum velocity in $5s$ duration after each lane change event.
 
\begin{figure}[h]
\renewcommand{\captionfont}{\small}
\centering\includegraphics[width=0.86\linewidth]{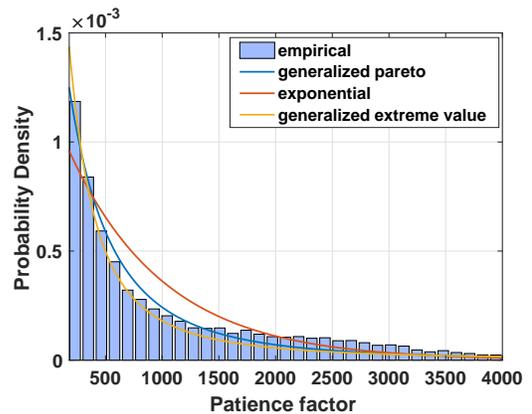}
\caption{Probability density distribution of the patience factor $\alpha_{pa}$.}
\label{fig:task1s2:patientdistribution}
\end{figure}

To calibrate the patience factor $\alpha_{pa}$, another 85,000 overtaking events are qualified out from SPMD database. 
Figure~\ref{fig:task1s2:patientdistribution} shows the probability density distribution of the patience factor in urban driving condition. The distribution can be fitted approximately by specific distribution functions such as generalized pareto distribution, generalized extreme value distribution and exponential distribution.

\subsection{Car following model}
Combining the politeness model and patience model as mentioned above, the lane change behavior of the human driver can be simulated. Since the model is formulated based on acceleration, an underlying longitudinal driving model needs to be specified. In this paper, we applied the errable modeling concept to simulate the human vehicle longitudinal driving behavior. 
For more details, please refer to~\cite{Yang2010}.

\section{The CAV driving trajectory generation}\label{section:CAVmodel}
In this section, in order to realize the interaction between human controlled vehicles and CAVs in simulation,
a standard Eco-AD algorithm~\cite{Guo2016,Kamal2011} is implemented to generate the speed trajectory of the CAV without considering cutting-in vehicles. Additionally, the  Adaptive Cruise Control (ACC) and Autonomous Emergency Braking (AEB) are considered to adjust the CAV's speed to maintain a safe distance towards the cutting-in vehicles.
\subsection{Eco-AD algorithm}
\begin{figure}[htp!]
\renewcommand{\captionfont}{\small}
\centering\includegraphics[width=0.92\linewidth]{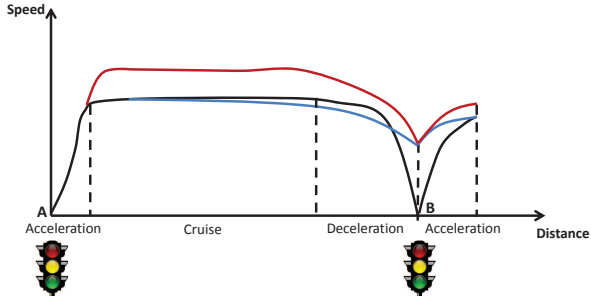}
\caption{Illustration of different driving trajectories of CAV in single traffic light scenario.}
\label{fig:task1s2:singleintersection}
\end{figure}

As shown in Fig.~\ref{fig:task1s2:singleintersection}, 
the objective of Eco-AD is to find the optimum vehicle velocity trajectory based on the knowledge of traffic signals schedule and map information between intersection A and intersection B which achieves the lowest energy consumption. 

The Eco-AD can be formulated by solving an optimization problem as,
\begin{equation}
\label{eq:task1s2:optimalformulation}
minJ=\varphi(x(t_f))+\int_{t_0}^{t_f} L(x, u, t)dt, ~~t\in[t_0, t_f],
\end{equation}
s.t.
\begin{equation}
\label{eq:task1s2:optimalformulationconstraint}
\begin{array}{l}
\dot{V}_{cav}(t) =\frac{1}{M}(F_t(t)-F_b(t)-\frac{1}{2}\rho c_d A_f V^2_{cav}(t) )-c_{r}g, \\
\dot{s}(t) =V_{cav}(t), \\
 \varphi(x(t_f))=(V_{cav}(t_f)-V_f)^2+(s(t_f)-s_f)^2,\\
 V_{cav}(t_0) =v_0, ~~s(t_0)=s_0, \\
 v_{min}\leq V_{cav}(t) \leq v_{max},  \\
 F_{t,min}\leq F_t(t) \leq F_{t,max},    \\
 0 \leq F_b(t) \leq F_{b,max},    \\
\end{array}
\end{equation}
where the state variable $x=[V_{cav},s]^T$ includes the CAV speed $V_{cav}$ and travel distance $s$, the control variable $u=[F_t(t), F_b(t)]^T$ includes the driving force $F_t(t)$ and braking force $F_b(t)$, $t_0$ denotes the initial value of time, $t_f$, $V_f$, $s_f$ denote the final values of time, velocity and position respectively, $M$ is the equivalent mass of the vehicle, $\frac{1}{2}\rho c_d A_f$ is the aerodynamic drag resistance, and $c_rg$ represents the acceleration caused by rolling resistance and gradient resistance.

The cost function $L$ is given by
\begin{equation}
\label{eq:task1s2:costL}
L(x,u,t)=W_1 P_{loss}+W_2F^2_b,
\end{equation}
where $W_1, W_2$ are the weighting parameters on power loss and braking force. The power loss $P_{loss}$ is mainly depended on the driving resistance given by 
\begin{equation}
\label{eq:task1s2:ploss}
P_{loss}(x,u,t)=\frac{1}{2}\rho c_d A_f V^3_{cav}(t)+Mc_rgV_{cav}(t).
\end{equation}
Thus, based on (\ref{eq:task1s2:optimalformulation}) and (\ref{eq:task1s2:optimalformulationconstraint}), different Eco-driving trajectories can be generated by selecting different boundary values of $t_f$, $V_f$ and $s_f$.


\subsection{CAV longitudinal control}
When the human vehicles cut in, Eco-AD driving mode of CAVs will be terminated. Then the ACC is active to adjust the vehicle speed to maintain a safe distance from cutting-in vehicles and the AEB will become active to stop the vehicle if a fatal threat is detected~\cite{Zhao2016}.

For AEB algorithm, the collision risk is assessed by a threshold value of 'Time-To-Collision', defined by
\begin{equation}
\label{eq:task1s2:AEB}
TTC=-\frac{R_L}{\dot{R}_L}<TTC_{AEB},
\end{equation}
where $R_L$ is the relative range to the front vehicle and $TTC_{AEB}$ is the threshold to activate AEB.

For ACC algorithm,  a PI controller is used to adjust the vehicle speed to achieve the desired time headway $T^{ACC}_{HWd}$ as
\begin{equation}
\label{eq:task1s2:ACC}
\begin{array}{l}
a_d(t)= K_p e_{HW}+K_i\int ^{t}_0 e_{HW}(\tau)d\tau, \\
e_{HW}=t_{HW}-T^{ACC}_{HWd},\\
t_{HW}=R_L/V_{cav},
 \end{array}
\end{equation}
where $t_{HW}$ is the time headway, $K_p$ and $K_i$ are the control gains, $a_d$ is the acceleration command.
\section{Simulation and energy evaluation}
\label{section:simulation}
In this section, the proposed lane changing model of the human driver is applied in cut-in scenario. The sensitivity on CAV driving strategy and human driver's lane change behavior is investigated by simulation and the corresponding energy consumption of CAV is evaluated.

\subsection{Simulation for sensitivity analysis}
A case study on simulation is conducted in city driving between two intersections (A and B) with 0.6 $km$ distance of double-lane road involving human controlled vehicles and CAVs. In this simulation, we assume a CAV starts driving from intersection A without any preceding vehicles when the signal switches to green. A human controlled vehicle follows the CAV at this initial point. The politeness factor and the patience factor of the human driver are selected as $p=1$ and $\alpha_{pa}=500$ respectively.  

\begin{figure}[h]
\renewcommand{\captionfont}{\small}
\centering\includegraphics[width=0.95\linewidth]{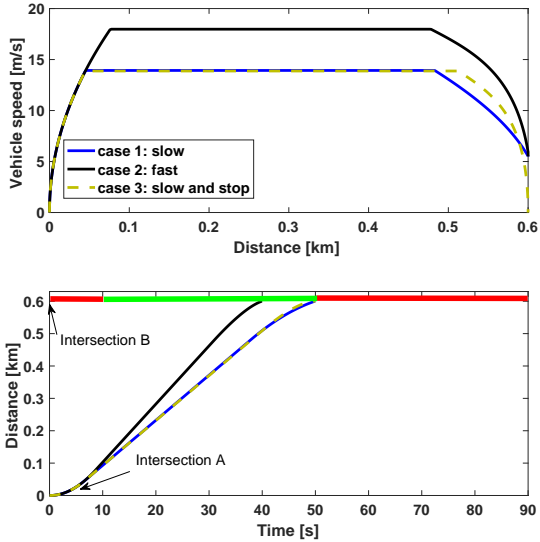}
\caption{Eco-driving trajectories in different cases: velocity trajectories v.s distance, distance trajectories v.s time. }
\label{fig:task1s2:drivingcases}
\end{figure}

Different trajectories of the CAV are considered and generated under different boundary conditions: 
\begin{itemize}
  \item Case 1: Traveling with slow cruising speed to pass the intersection B right before the next red phase;
  \item Case 2: Traveling with high cruising speed to pass the intersection B within the current green phase;
  \item Case 3: Traveling with the same cruising speed of Case 1 but stopping at the intersection B right as the red phase starts.
\end{itemize}
The driving trajectories of above 3 cases and the traffic signal timing are shown in Fig.~\ref{fig:task1s2:drivingcases}. 
Note that a relatively low terminal speed is selected to allow the CAV to slow down at intersection B for safety concerns.   


\begin{figure}[h]
\renewcommand{\captionfont}{\small}
\centering\includegraphics[width=1\linewidth]{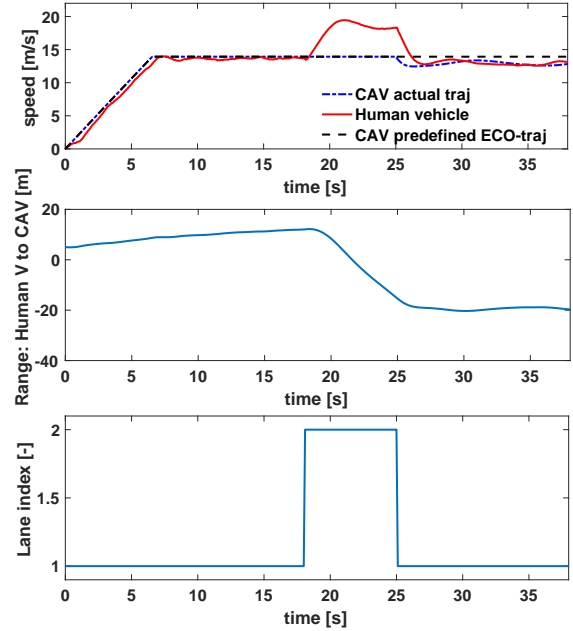}
\caption{Simulation result: the CAV travels with the slow cruising speed in Case 1. }
\label{fig:task1s2:case1}
\end{figure}

Fig.~\ref{fig:task1s2:case1} shows the human driver's behavior in Case 1 in which the CAV travels with slow cruising speed around $14m/s$.
At the beginning, the human vehicle follows the slow CAV for a while in the original lane. When the velocity loss exceeds the driver's patient tolerance and the traffic condition satisfies the politeness criterion, the human vehicle makes a lane change and then drives faster in the target lane. After passing the CAV, the human vehicle changes back to the original lane when the range to the CAV is safe. At this moment, the CAV switches the driving mode from Eco-driving to adaptive cruising to keep a safe time headway towards the preceding human vehicle. As the human vehicle overtakes the CAV and disturbs its Eco-driving trajectory, the CAV loses the chance to pass the intersection B within the current green phase. Consequently, the CAV has to stop at the intersection B until the next green signal starts (40s waiting time at the intersection B is considered in this simulation). 

\begin{figure}[h]
\centering\includegraphics[width=1.0\linewidth]{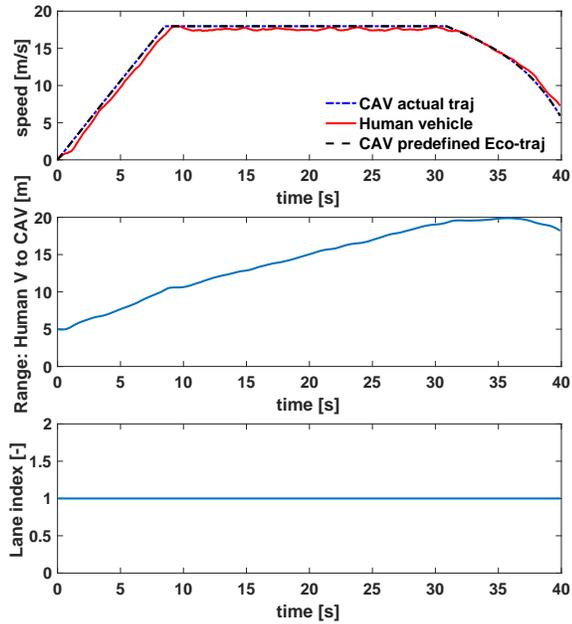}
\caption{Simulation result: the CAV travels with high cruising speed in Case 2.}
\label{fig:task1s2:case2}
\end{figure}

Fig.~\ref{fig:task1s2:case2} shows the human driver's behavior in Case 2 in which the CAV travels with a relatively high cruising speed (around $18m/s$). As the speed of CAV is close to human driver's desired speed, the human vehicle keeps following the CAV without losing patience and the CAV keeps driving in Eco-driving mode to pass the intersection B as expected. 
\subsection{Preliminary energy evaluation}
To evaluate the fuel consumption of the CAV in the above cases, we assume the CAV is driven by conventional internal combustion engine. The fuel consumption $f_{total}$ is approximately calculated by a polynomial equation depending on instantaneous vehicle velocity $V_{cav}$~\cite{Kamal2011},

\begin{equation}\label{eq:task1s2:energyconsumpmodel}
 f_{total}=f(V_{cav},V_{cav}^2, V_{cav}^3, \dot{V}_{cav}).
\end{equation}

\begin{figure}[h]
\renewcommand{\captionfont}{\small}
\centering\includegraphics[width=0.62\linewidth]{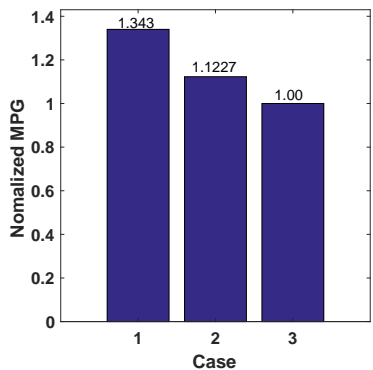}
\caption{Fuel economy evaluation of the CAV in the case study.}
\label{fig:task1s2:evaluation}
\end{figure}

The preliminary fuel economy evaluation of the simulation is shown in Fig.~\ref{fig:task1s2:evaluation}. As expected, the trajectory of the CAV in Case 1 leads to the best fuel economy when there is no cut-in vehicle; However, if human drivers happen to cut in, the CAV will lose the chance to pass the intersection B within the current green signal and the corresponding fuel economy will drop significantly. 
In such a case, the trajectory of CAV in Case 3 is used as the ideal case to evaluate the energy consumption.
For the trajectory of the CAV in Case 2, more fuel is consumed comparing with Case 1 as a higher average speed is executed. But the higher speed will prevent the surrounding vehicles from cutting in and make the CAV pass the intersection B as expected. Overall, over $12\%$ of fuel economy can be improved comparing with the stop case. Note that the fuel economy evaluation depends on the red signal timing and the propulsion type of the CAV, future work will focus on a comprehensive simulation and evaluation in multi-traffic scenarios.


\section{Conclusion}\label{section:conclusion}
This paper presents the evaluation of energy efficiency for CAVs in urban driving condition based on a microscopic lane changing model. Particularly, a stochastic lane changing model emulating the characteristics of politeness and patience of human driver is developed based on naturalistic database. 
The proposed lane changing model is applied in the cut-in scenario to simulate the interaction between human controlled vehicles and CAVs. A case study on simulation is conducted under different driving trajectories of a CAV near a single intersection and the corresponding impact on the fuel consumption of the CAV is preliminarily evaluated. The result shows the potential of the improved fuel economy for the CAV if the cut-in vehicle can be appropriately handled. Future work will focus on the simulation and evaluation in the multi-traffic scenarios.

%
%
%
%


\bibliographystyle{IEEEtran}
\bibliography{ref.bib}

\end{document}